\documentclass[aps, reprint, onecolumn, notitlepage, prl,   amssymb, amsmath, superscriptaddress, showpacs] {revtex4-1}
\usepackage{graphicx}
\usepackage{hyperref}
\usepackage{bm}
\usepackage{enumitem}
\usepackage{subfigure}

\begin{document}
	
	\title{Ultrafast magnetization dynamics in uniaxial ferrimagnets with compensation point. GdFeCo} 

	\begin{abstract}
		
		We derive an effective Lagrangian in the quasi-antiferromagnetic approximation that allows to describe the magnetization dynamics for uniaxial $f$-$d$ (rare-earth - transition metal) ferrimagnet near the magnetization compensation point in the presence of external magnetic field. We perform calculations for the parameters of GdFeCo, a metallic ferrimagnet with compensation point that is one of the most promising materials in  ultrafast magnetism. Using the developed approach, we find the torque that acts on the magnetization due to ultrafast demagnetization pulse that can be caused either by ultrashort laser or  electrical current pulse. 
		We show that the torque is non-zero only in the non-collinear magnetic phase that can be acquired by applying external magnetic field to the material. The coherent response of magnetization dynamics amplitude and its timescale exhibits critical behavior near certain values of the magnetic field corresponding to a spin-flop like phase transition.  
		Understanding the underlying mechanisms for these effects opens the way to efficient control of the amplitude and the timescales of the spin dynamics, which is one of the central problems in the field of ultrafast magnetism. 
	\end{abstract} 	
	
	\pacs{79.20.Ds, 75.50.Gg, 75.30.Kz, 75.78.-n}
	
	\author{M.~D.~Davydova}
	\email{davydova@phystech.edu}
	\affiliation{Prokhorov General Physics Institute of the Russian Academy of Sciences, 119991 Moscow, Russia}
	\affiliation{Moscow Institute of Physics and Technology~(State University), 141700 Dolgoprudny, Russia}
	
	\author{K.~A.~Zvezdin}
	\affiliation{Prokhorov General Physics Institute of the Russian Academy of Sciences, 119991 Moscow, Russia}
	\affiliation{Moscow Institute of Physics and Technology~(State University), 141700 Dolgoprudny, Russia}
	
	\author{A.~V.~Kimel}
	\affiliation{Moscow Technological University (MIREA), 119454 Moscow, Russia}
	\affiliation{Institute for Molecules and Materials, Radboud University, Nijmegen  6525 AJ, The Netherlands}
	
	\author{A.~K.~Zvezdin}
		\email{zvezdin@gmail.com}
	\affiliation{Prokhorov General Physics Institute of the Russian Academy of Sciences, 119991 Moscow, Russia}
	\affiliation{Moscow Institute of Physics and Technology~(State University), 141700 Dolgoprudny, Russia}
	
	\maketitle

	\section{Introduction}

	Most of the prominent advances in the field of ultrafast magnetism  have been achieved by using thermal  mechanism of magnetization control \cite{ostler2012ultrafast,radu2011transient,PhysRevLett.99.047601,PhysRevB.88.020406,PhysRevLett.108.127205,PhysRevB.96.014409,PhysRevB.87.224417}. These studies rooted from the pioneering work by Beaurepaire et al. \cite{PhysRevLett.76.4250} on ultrafast laser-induced demagnetization of Ni. In this experiment, partial destruction of magnetic order was found at much faster rates that were believed to be possible prior to that publication. Since then, the field of ultrafast magnetism has been rapidly growing and the possible channels of ultrafast angular momentum transfer have been studied extensively \cite{RevModPhys.82.2731}. 	Ultrafast demagnetization can be achieved by applying ultrashort laser pulses \cite{ostler2012ultrafast,radu2011transient,PhysRevLett.99.047601,PhysRevB.88.020406,PhysRevLett.108.127205,PhysRevB.87.224417,PhysRevB.96.014409}, or, alternatively, by using short pulses of  electric currents \cite{PhysRevB.95.180409,Yange1603117}. 
	
	In the last decades, GdFeCo and other rare-earth - transition metal compounds (RE-TM) have been in the center of attention in this regard \cite{PhysRevB.92.094411}. For example, all-optical switching has been demonstrated for the first time in GdFeCo \cite{PhysRevLett.99.217204}. It was found that the switching is possible due to different rates of sublattice demagnetization, which enables  ultrafast magnetization reversal to occur because of the angular momentum conservation \cite{radu2011transient,ostler2012ultrafast}.
	
	In many RE-TM compounds, GdFeCo and TbFeCo being part of them, realization of the magnetization compensation point is possible. At this point, the magnetizations of the two antiferromagnetically coupled sublattices with different dependencies on temperature become equal and the total magnetization of the material turns to zero. In the presence of the external magnetic field, a record-breaking fast subpicosecond magnetization switching was found in GdFeCo across the compensation point \cite{PhysRevLett.99.217204}. In addition, a number of anomalies in the magnetic response was observed near this point \cite{PhysRevB.73.220402,PhysRevLett.118.117203,doi:10.1063/1.3462429}, which has never been explained theoretically. All said above illustrates the importance of understanding  the role of the compensation point in the dynamics and working out an appropriate tool for its description. 
		 
	Efficient control of the amplitude and the timescales of the response of the magnetic system to an ultrafast demagnetizing impact on a medium is one of the most important issues in the area of ultrafast magnetism nowadays \cite{RevModPhys.82.2731, PhysRevB.73.220402, PhysRevLett.118.117203}. Understanding of the mechanisms and of the exhaustive description of the subsequent spin dynamics is also a long-standing goal that will help to promote the achievements of this area towards practical applications in magnetic recording \cite{PhysRevLett.99.047601,moser2002magnetic}, magnonics \cite{LENK2011107} and spintronics \cite{doi:10.1063/1.4958846}. In this work, we expand the understanding of response of magnetic system of a uniaxial $f$-$d$ ferrimagnet  near the compensation point in the external magnetic field to an ultrafast demagnetizing pulse, which can be induced either by a  femtosecond laser or an electric current pulse. We present a theoretical model and calculations, which allow to describe the ultrafast response of the system that resides in an angular phase before the impact. We show that in this magnetic phase the coherent precessional response is possible  and the subsequent magnetization dynamics may become greatly nonlinear and is governed by large intersublattice exchange field \cite{PhysRevB.46.11201}. We derive the effective Lagrangian that governs the dynamics of the system near the compensation point and obtain the torque acting on the magnetizations of the two sublattices due to demagnetization. 
	In ref. \cite{PhysRevLett.118.117203}, the critical response of the amplitude and the time of the signal rise have been found in GdFeCo in external magnetic field along the easy axis. At given laser pump fluences, the response was found to be negligible in collinear phase, but it was dramatically large in angular one.  We elaborate on this example and show that the critical behavior of the response is the consequence of the second-order magnetic phase transition from collinear to an angular in the external magnetic field. We find that these effects are pronounced in the vicinity of the compensation point, where the phase transitions cross each other\cite{goransky1970temperature,zvezdin1972some,sabdenov2017magnetic1}. Thus, the proposed model explains a range of important experimental observations as well as allows for developments of methods and tools of magnetization control by setting the temperature near the compensation point and applying magnetic field. Moreover, by changing the composition of the alloy, the \cite{DING201351}, the position of the magnetization compensation point can be tuned arbitrary close to the room temperature. Our results might open new ways for  technologies for ultrafast optical magnetic memory.

	\section{Effective Lagrangian and Rayleigh disspation function}	
	
	Our approach is based on Landau-Lifshitz-Gilbert equations for a two-sublattice (RE-TM) ferrimagnet. These equations are equivalent to the following effective Lagrangian and Rayleigh dissipation functions:
	\begin{equation} \label{lagr0}
	\mathcal L = \frac{M_f}{\gamma}	 \left( 1 - \cos \theta_f\right) \frac{\partial \varphi_f}{\partial t} + \frac{M_d}{\gamma}	 \left( 1 - \cos \theta_d\right) \frac{\partial \varphi_d}{\partial t} - \Phi(\bm M_f, \bm M_d, \bm H),
	\end{equation}
	\begin{equation} \label {R0}
		\mathcal R=	\mathcal R_{f}+	\mathcal R_{d}, \ \ \mathcal R_{f,d} =   \frac{\alpha M _{f,d}}{2 \gamma} \left(\dot \theta_{f,d}^2 + \sin^2 \theta_{f,d} \dot \varphi_{f,d}^2 \right) 
	\end{equation}
	where $\gamma$ is the gyromagnetic ratio,  $\bm M_d$ and $\bm M_f$ are the magnetizations, $\theta_d$ (TM) and $\theta_f$ (RE) are the polar, $\varphi_d$ and $\varphi_f$ are the azimuthal angles of $d$- and $f$- sublattices correspondingly in the spherical system of coordinates with $z$-axis aligned along the external magnetic field $\bm H$.   $\Phi(\bm M_f, \bm M_d, \bm H)$ is the thermodynamic potential for the system that we take in the following form:
	\begin{equation}
	\Phi = -\bm M_d \bm H + \lambda \bm M_d \bm M_f  -\bm M_f \bm H - K_f \frac{( \bm M_f \bm n)^2 }{M_f^2}- K_{d} \frac{( \bm M_d \bm n)^2 }{M_d^2},	
	\end{equation}
	where  $\lambda$ is the intersublattice exchange constant, $\bm n$ is the direction of the easy axis and $K_{f,d}$ are the anisotropy constants for $f$- and $d$- sublattices, respectively.

	Next, we transfer to description in terms of the antiferromagnetic $\bm L = \bm M_R- \bm M_d$ and the total magnetization $\bm M = \bm M_R + \bm M_d$ vectors. 	In the vicinity of the compensation point the difference between the sublattice magnetizations $|M_R-M_d |\ll L$ is small.
 	The two vectors are parametrized using the sets of angles $\theta, \varepsilon$  and $\varphi, \beta$, which are defined as:
	\begin{equation} \label{angles}
	\begin{split}
	\theta_f &= \theta -\varepsilon, \ \theta_d = \pi - \theta - \varepsilon, \\
	\varphi_f &= \varphi + \beta, \ \varphi_d = \pi + \varphi - \beta.
	\end{split}	
	\end{equation}
	In this case the antiferromagnetic vector is naturally defined as $\bm L =  (L \sin \theta \cos \varphi, L \sin \theta \sin \varphi, L \cos \theta)$. 
	
	Following the work \cite{zvezdin1979,*zvezdin2017dynamics} we use the quasi-antiferromagnetic approximation to describe the dynamics near the magnetization compensation point.  F
	In this approximation the canting angles are small $\varepsilon\ll 1$, $\beta \ll 1$, and we can expand the Lagrangian \eqref{lagr0} and the corresponding thermodynamic potential up to quadratic terms in small variables:
	\begin{equation} \label {L}
	\begin{split}
	\mathcal L &= -\frac{m}{\gamma}\dot \varphi \cos \theta - \frac{M}{\gamma} \sin \theta \left(\dot \varphi \varepsilon - \beta \dot \theta  \right) -\Phi, \\
	\Phi &= - K (\bm l, \bm n)^2 - H m \cos \theta - \varepsilon M H \sin \theta + \frac{\delta}{2} \left( \varepsilon^2 + \sin^2 \theta \beta^2 \right).
	\end{split}
	\end{equation}
	Here $m = M_R - M_d$, $M = M_R + M_d$, $K = K_R + K_d$ is the effective uniaxial anisotropy constant, $\bm l = \bm L / L$ is the unit antiferromagnetic vector $\delta = 2 \lambda M_d M_R$ and we assume the anisotropy to be weak $K\ll  \lambda M$. For GdFeCo with 24\% Gd and compensation point near 283 K, we assume the following values of parameters: $M \approx 800$ emu/cc, $K = 1.5 \times 10^5$ erg/cc, $\lambda = 18.5$ T/$\mu_B$, $\delta \approx  10^9$ erg/cc and $m$ changes in the range between $150$ emu/cc and $-50$ emu/cc at fields $H \approx H^* \approx 4$ T. The characteristic values of small angles $\epsilon$ and $\beta$ are of the order of $10^{-2}$.
	
	Next, we exclude the variables $\varepsilon$ and $\beta$ by solving the Euler-Lagrange equations.
	Substituting them into the Lagrangian \eqref{L}, we obtain the effective Lagrangian, which describes the dynamics of a uniaxial ferrimagnet in the vicinity of the compensation point:
	\begin{equation} \label {L1}
	\mathcal L_{eff} =   \frac{\chi_\perp}{2} \left( \frac{\dot \theta}{\gamma}\right) ^2 + m \cos \theta \left( H - \frac{\dot \varphi}{\gamma}\right) + \frac{\chi_\perp}{2} \sin^2 \theta  \left( H - \frac{\dot \varphi}{\gamma}\right)^2    + K (\bm l, \bm n)^2,
	\end{equation}
	\begin{equation} \label {F1}
	 \Phi_{eff}(H) =    - m H \cos \theta    - \frac{\chi_\perp}{2} H ^2 \sin^2 \theta     - K (\bm l, \bm n)^2,
	\end{equation}
	\begin{equation} \label {R1}
	\mathcal R_{eff} =   \frac{\alpha M}{2 \gamma} \left(\dot \theta^2 + \sin^2 \theta \dot \varphi^2 \right) 
	\end{equation}
	where $\chi = \frac{2 M^2}{\delta}$. In GdFeCo $\chi \approx 1.6 \times 10^{-3}$ and $\alpha \approx 0.05$. 
	In the derivation above we assumed the gyrotropic factor $\gamma$ and Gilbert damping constant $\alpha$ to be the equal for both sublattices. Taking into account the difference between these values for different sublattices will lead to the angular momentum compensation effect at certain temperature. The Lagrangian, Rayleigh function and  equations of motion preserve the same form in this case if we substitute the parameters $\gamma$ and $\alpha$ with temperature-dependent factors $\widetilde \gamma$ and $\widetilde \alpha$ defined as:
	
		\begin{equation}
	 \frac{1}{\widetilde \gamma} = \frac{1}{\bar \gamma}\left(1 + \frac{M}{m} \frac{\gamma_f - \gamma_d }{\gamma_f+ \gamma_d} \right) =\frac{\frac{M_d}{\gamma_d} - \frac{M_f}{\gamma_f}}{(M_d - M_f)},\  \frac{1}{\bar \gamma} = \frac{1}{2}\left( \frac{1}{\gamma_d}+\frac{1}{\gamma_f}\right) , \
	\widetilde \alpha = \frac{\left(\alpha_d \gamma_f + \alpha_f \gamma_d \right) }{(\gamma_f + \gamma_d)}\frac{1}{1 + \frac{M}{m}\frac{\gamma_f - \gamma_d}{\gamma_f + \gamma_d}}
	\end{equation}
	
	This allows to reproduce the angular moment compensation phenomenon, which was studied experimentally in ref. \cite{PhysRevB.73.220402}.

	\section{Excitation of the spin dynamics  }	
	
	The proposed approach presents a powerful tool allowing analyzing coherent magnetization dynamics in ferrimagnets that occurs under a broad range of conditions. Let us consider the following example that poses an important problem in the field of ultrafast magnetism. An femtosecond laser pulse strikes the uniaxial ferrimagnet (for instance, of GdFeCo, TbFeCo type) in the presence of external static magnetic field. The impact of the laser pulse leads to the demagnetization of one or both of the sublattices. What coherent magnetization dynamics will occur as a consequence of this impact? The proposed model can be further developed in order to answer to this question and is applicable for small values of demagnetization $\delta M$. 
	
	In our framework the spin dynamics in ferrimagnet is described by Euler-Lagrange equations of the form $\frac{\mathrm{d}}{\mathrm{d}t}\frac{\partial \mathcal L}{ \partial \dot q} - \frac{\partial \mathcal L}{ \partial  q} = - \frac{\partial \mathcal R}{ \partial \dot q}$, where $q= \theta, \ \varphi$ are the polar and azimuthal angles describing the orientation of the antiferromagnetic vector $\bm L$, correspondingly.  Let us consider a particular case when the easy magnetization axis is aligned with the external magnetic field, which leads to the presence of azimuthal symmetry in the system.  In this case $\bm n = (0,0,1)$.
	In this particular case the Euler-Lagrange equations can be rewritten as:
	\begin{equation} \label {el1}
	\frac{\chi_\perp}{\gamma^2} \ddot \theta = \frac{\partial \mathcal{L}_{eff}}{\partial \theta} - \frac{\partial \mathcal{R}_{eff}}{\partial \dot \theta}, \
	\frac{d }{ dt } \frac{\partial \mathcal{L}_{eff}}{\partial \dot \varphi} =  - \frac{\partial \mathcal{R}_{eff}}{\partial \dot \varphi}
	\end{equation}
	
	The nonlinear equations that are similar to Eqs. \eqref{el1} and describe the spin dynamics of  two-sublattice ferrimagnets were obtained in the work \cite{ivanov1983nonlinear} under the conditions  $H = 0 $ and $\mathcal R_{eff} = 0$. Over the short time of demagnetization the second equation can be approximately treated as a conservation law and the conserving quantity (angular momentum of magnetization precession $\mathcal J$) stays approximately constant as $\partial \mathcal L / \partial \varphi = 0$ due to the Noether theorem:
	
	\begin{equation} \label {J}
	\mathcal J = \frac{\partial \mathcal{L}_{eff}}{\partial \dot \varphi} = -\frac{1}{\gamma} \left[ m \cos \theta + \chi_\perp \sin^2 \theta \left(H - \frac{\dot \varphi}{\gamma} \right)  \right] = const
	\end{equation}

	Let the moment of time $t= 0-$ denote the moment before the laser pulse impact and system initially is in the ground state defined by the ground state angles $\theta(0-) = \theta_0$, $\varphi (0-) = \varphi_0$, and their derivatives $\dot \varphi(0-) = 0$, $\dot \theta(0-) = 0$. Depending on the external parameters and preparation of the sample, the system might reside in one of the two possible antiferromagnetic collinear phases or in angular phase, which are separated by the magnetic phase transition lines \cite{zvezdin1995field}. If the demagnetization due to the laser pulse action is small, it produces the changes in the values of $M_f$, $M_d$ and $M$ of the order of percent or less, whereas the change of $m$ (which is approximately equal to total magnetization near the compensation point) may be of several orders of magnitude, as its value is almost compensated. In what follows, we assume that the demagnetization is associated only with change of $m$, namely  $m = m_0 + \Delta m(t)$. As we will see below, the change in this quantity already leads to several drastic effect in dynamics. 

	Therefore, the conservation law \eqref{J} leads to the emergence of azimuthal dynamics $\dot \varphi(t)$ at the demagnetization timescales ($\Delta t$) due to demagnetization pulse $\Delta m(t)$:
		\begin{equation} \label {phidot}
	\dot \varphi(t) = \gamma \frac{\Delta m(t)}{ \chi_\perp} \frac{\cos \theta_0}{\sin^2 \theta_0}
	\end{equation}

	We see that the torque is non-zero only in the angular phase, where $0< \theta_0 < \pi$. Emergence of the azimuthal spin precession as a result of demagnetization of the medium is similar to the well-known Einstein-de-Haas effect, where the demagnetization leads to azimuthal precession of the body.  Subsequently, this azimuthal spin dynamics leads to the emergence of polar dynamics $\theta(t)$, which is most commonly measured  in pump-probe experiments of ultrafast magnetism, by acting as an effective field $H_{eff} = H - \frac{\dot \varphi}{\gamma}$ in the Lagrangian \eqref{L}. We can then view the Lagrangian as depending only on variable $\theta$ and the effective field $H_{eff}$. At demagnetization $\delta m \sim 0.01 M$ in GdFeCo the value of $\dot \varphi$ can reach up to 1 THz, and the corresponding effective magnetic field is of the order of 10 T. Note that initial state of the system corresponds to the condition $\frac{\partial \Phi}{ \partial \theta_0}(H_{eff} = H) = 0$.
	 We can rewrite the Euler-Lagrange equation from eq. \eqref{el1} for polar angle as follows: 
		\begin{equation} \label {el11}
		\frac{\chi_\perp}{\gamma^2} \ddot \theta + \frac{\partial \Phi(H_{eff})}{\partial \theta} =- \frac{\alpha M}{\gamma} \dot \theta.
		\end{equation}
		
		Or, alternatively:
			\begin{equation} \label {el12}
			\frac{\chi_\perp}{\gamma^2} \ddot \theta + m \sin \theta H_{eff} - \chi_\perp \sin \theta \cos \theta \left(H_{eff}^2 - \frac{2 K}{\chi_\perp} \right)  =- \frac{\alpha M}{\gamma} \dot \theta
			\end{equation}

	By integrating this equation over the short demagnetization pulse duration $\Delta t$  we obtain the state of system after the laser pulse impact at $t = 0+$, which is characterized by the initial conditions
	\begin{equation*}
	 \theta (0+) = \theta_0, \ \dot \varphi (0+), \ \varphi (0+) = \int_0^{\Delta t} \dot \varphi(t) \mathrm{d} t, \ \dot \theta (0+) = \int _0^{\Delta t} \ddot \theta (t) \mathrm {d} t 
	\end{equation*}
	The value  $\Delta t$ is of the order of the optical pulse length. It may also include the time of restoration of the magnetization length (or the value of  $m$). After the moment of time $0+$ free magnetization precession occurs in the model. Analysis of the spin dynamics under laser pump excitation will lead to emergence of critical dynamics near the second-order phase transitions to the collinear phases where $\theta = 0,\pi$, as is already seen from \eqref{phidot}. We will discuss this behavior below.

				\begin{figure}[htb]%
				\includegraphics[width=0.4\columnwidth]{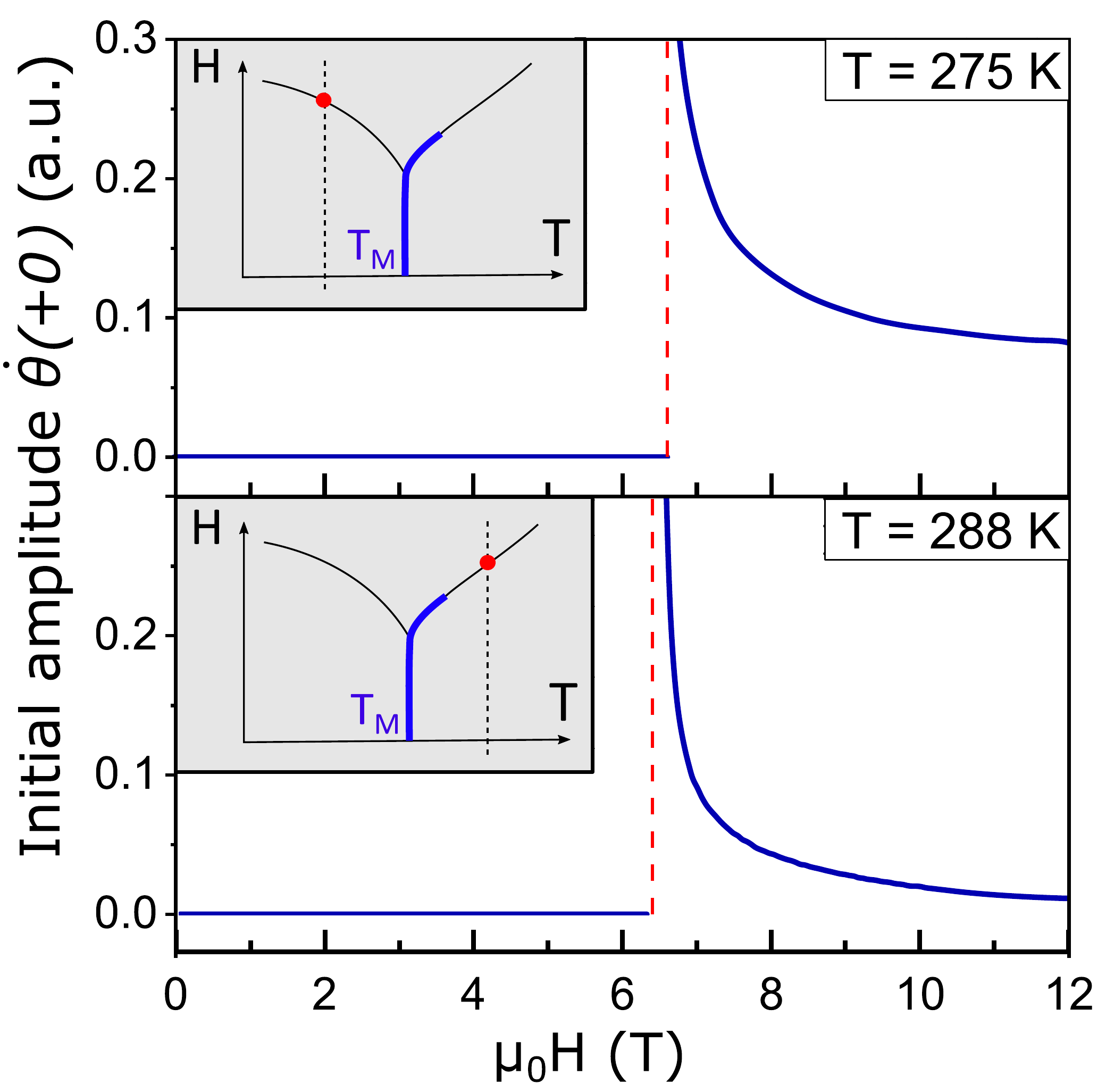}%
				\caption{%
					The amplitude of  the magnetization precessional response  after the demagnetization due to the femtosecond laser pulse action in GdFeCo ferrimagnet near the compensation point at different temperatures.	Insertions: the schematic of the magnetic phase diagram. There are two antiferromagnetic collinear phases with $\bm M_d$ directed along(opposite) to the external magnetic field above(below) the compensation temperature $T_M$. They are separated by the first-order phase transition line (blue). Above them, an area where the angular phase exists, which is filled with gray color. The black solid lines  are the second-order phase transition lines. The dashed lines corresponds to the fixed temperature in the plot. The red dot is the point of phase transition for this temperature. 	}
				\label{ampl}
			\end{figure}
			
	\section{Critical dynamics}	
			
		In a simple case of a quick decay of demagnetization (at the exciton relaxation timescales) with $\Delta m (t) = \Delta m$, $0<t<\Delta t$; $\Delta m(t) = 0$, $t>\Delta t$, we obtain  the initial condition from \eqref{el12}:	
	
	\begin{equation} \label{angle}
	\begin{split}
	\dot \theta(+0) \approx \left[- \left( 2 \frac{\cos^2 \theta_0}{\sin \theta_0} + \sin \theta_0 \right) H +  \frac{m_0}{\chi_\perp} \frac{\cos \theta_0}{\sin \theta_0}+\frac{\Delta m}{\chi_\perp} \frac{\cos \theta_0}{\sin^3 \theta_0} \right] \frac{\gamma^2 }{\chi_\perp} \Delta m \Delta t = B(\theta_0) \Delta m + O(\Delta m^2).
	\end{split}
	\end{equation}
			
	\noindent This quantity defines the initial angular momentum of the polar spin precession that is induced in the system due to the optical spin torque created by the femtosecond laser pulse.  The amplitude of oscillations is proportional to the initial condition \eqref{angle}. Its dependence on the external magnetic field is illustrated in Fig. \ref{ampl} for different temperatures for magnetic parameters of GdFeCo uniaxial ferrimagnet. At low values of external magnetic fields there is only collinear ground state in the ferrimagnet and above certain field $H_{sf}$ the transition to an angular state occurs \cite{goransky1970temperature}. The schematic of the magnetic phase diagram for GdFeCo is shown in insertions in Fig. \ref{ampl}. 
	 At $T=275$ K and $T=288$ K the phase transitions are of the second order, which corresponds to a smooth  transition from angle $\theta_0 = 0$ to $\theta_0  > 0 $, and the divergence of the response occurs at $H_{sf}$.  Immediately above the compensation temperature the transition is of the first order and the behavior of the response above the  is more complex; however, there is no critical divergence. The critical behavior of the signal amplitude was observed experimentally for GdFeCo  in ref. [\onlinecite{PhysRevLett.118.117203}].

	Another feature in the dynamics described by the proposed model is the critical behavior of the characteristic timescales that occurs in the vicinity of the second-order phase transitions. To demonstrate this effect analytically, we assume small deviations of $\theta$ during oscillations: $\theta(t) = \theta_0 + \delta \theta(t)$. We obtain:
	
	\begin{equation} \label{lin}
	\delta \ddot \theta + \omega_r^2(\theta_0) \delta \theta = - \alpha \omega_{ex} \delta \dot \theta,
	\end{equation}

	where $\omega_r^2(\theta_0) = \gamma^2\left[ \frac{m}{\chi_\perp} H \cos \theta_0 + \left(\frac{2 K}{\chi_\perp} -H^2\right) \cos 2 \theta_0 \right]$, $\omega_{ex} = \gamma \frac{M}{\chi_\perp}$. The initial conditions are $\theta(0) 	= \theta_0$ and eq. \eqref{angle}.
	In the limit of small oscillations and $\omega_r < \alpha \omega_{ex}/2$ (is fulfilled near the second-order transition) the solution has the form $\delta \theta(t) = A e^{-\beta t} \sinh \omega t$, where $\beta = \alpha \omega_{ex}/2$, $\omega^2 =  \beta^2 -\omega_r^2 $, $A = B(\theta_0)/\omega$. 
	The rise time can be estimated from the condition $\dot \theta(\tau_{rise}) = 0$:
	
	\begin{equation}
	\tau_{rise} \approx \frac{\mathrm{atanh} \frac{\omega}{\beta}}{\omega} = \frac{\mathrm{atanh} \frac{\sqrt{ \beta^2 - \omega_r^2 }}{\beta}}{\sqrt{\beta^2 - \omega_r^2}}
	\end{equation}		
		
	 The time of the oscillations decay (relaxation time) is proportional to the imaginary part of eigenfrequency and can be estimated by the following expression:
	
			\begin{equation}
			\tau_{relax} \approx \frac{4 \pi \beta}{\omega_r^2}.
			\end{equation}	
			
	Near second-order phase transition the mode softening occurs and the eigenfrequency turns to zero: $\omega_r \rightarrow 0$,  and we observe growth of the both timescales.  The critical behavior of the rise time has been observed in GdFeCo experimentally \cite{PhysRevLett.118.117203} and the typical values of $\tau_{rise}$ were of the order of 10 ps.

			\section*{Conclusions}

		To sum up, the developed theoretical model based on quasi-antiferromagnetic Lagrangian formalism proved to be suitable for description of the coherent ultrafast response of RE-TM ferrimagnets near the compensation point due to an ultrashort pulse of demagnetization in the presence of external magnetic field. We have  found that the torque acting on magnetizations is non-zero in the noncollinear phase only. We have explained  the experimentally observed critical behavior of the response amplitude and characteristic timescales as the consequence of the second-order magnetic phase transition from collinear to an angular in the external magnetic field and the mode softening near it. These effects are vivid in the vicinity of the compensation point in external magnetic field. Understanding the ultrafast response to demagnetizing optical or electrical pulses and subsequent spin dynamics can facilitate future developments in the fields of ultrafast energy-efficient magnetic recording, magnonics  and spintronics.

		\section*{Acknowledgments}
		
		This research has been supported by RSF grant No. 17-12-01333.

\end{document}